\documentclass{PoS}
\newcommand{\be}{\begin{eqnarray}}
\newcommand{\ee}{\end{eqnarray}}
\title{ELECTRIC-MAGNETIC STRUGGLE IN QGP, DECONFINEMENT AND BARYONS}

\ShortTitle{Electric-magnetic struggle...}

\author{\speaker{Edward Shuryak}\thanks{The talk is based on 
several works
        done in collaboration with J-F.Liao }\\
        Department of Physics and Astronomy, University at Stony Brook,\\
Stony Brook NY 11794 USA \\
        E-mail: \email{shuryak@tonic.physics.sunysb.edu}}

\abstract{ We argue that quite unusual properties
of Quark-Gluon Plasma in the RHIC temperature domain
$T=(1-2)T_c$ are consequences of the approximate equilibrium between
electric and magnetic sectors reached above the deconfinement temperature.
Already classical study of few body motion in a electric-magnetic 
plasma shows abnormally large scattering rate due to the so called
``magnetic bottle'' effect. Molecular dynamics simulation have found
that equal mixture of electric and magnetic quasiparticles do produce
plasmas of  small viscosity and even smaller diffusion rate, comparable
to what is needed to explain RHIC data and also to 
what follows from AdS/CFT.
 As a separate issue, we point out that right above $T_c$ there should
 be surviving s-wave baryons made of quarks ($N,\Delta$), as well as
adjoint objects, 3-gluon and 3-monopole chains (the latter being
nothing else but ``calorons'' or finite-T instantons).          }

\FullConference{Critical Point and Onset of Deconfinement
          4th International Workshop\\
		 July 9-13 2007\\
		 GSI Darmstadt,Germany}

\begin{document}

\section{Why is sQGP at RHIC such a perfect fluid?}
 A realization~\cite{Shu_liquid,SZ_rethinking,SZ_CFT} that
QGP at RHIC  is not a weakly coupled 
gas but rather a strongly coupled liquid
 has lead to a  paradigm
 shift in the field. It was extensively debated  at
the ``discovery''  BNL workshop in 2004~\cite{discovery_workshop} (at which the
abbreviation sQGP was established) and multiple other meetings since.

There is no need to repeat any of the arguments, except to note
key new observations. Experimental findings indicating that
charm quarks are as strongly quenched as light (gluonic) jets,
and participate in elliptic flow, have shown that not only viscosity
$\eta$
but heavy quark diffusion constant $D$ is anomalously small \cite{MT}.
  Further conformations of 
{\em conical} flow from quenched jets, suggested in 
\cite{Stoecker:2004qu,CST}
came  both from RHIC (3-particle
correlations from PHENIX and STAR) and from
impressive AdS/CFT calculations \cite{Friess:2006fk,Chesler:2007an}. All these facts
show that jet quenching mechanism is very different from
the perturbative picture of the
gluon emission: it is rather an emission of intense sound/wake waves 
as pointed  out in \cite{CST}.

In the intervening  years we
had to learn a lot, some new some 
 from other branches of physics  which
happened to have some experience with strongly coupled systems.
Those range from
quantum gases to classical plasmas to string theory.
In short, there seem to be not one
but  actually $two$ difficult
issues we are facing: \\
(i) to understand whether QGP is
(or is not) strongly
coupled in the quasi-conformal regime, at $T> 2T_c$ ;
 (ii) to understand physics near the
 deconfinement transition, at $T\approx T_c$.

 Comparison of the results of the AdS/CFT correspondence
-- adequate for quasi-conformal strongly coupled plasma --  to data
(see the end of section \ref{sec_MD}) is encouraging enough
to think that  the answer to (i) may be positive.
This created a lot of excitement, especially among string theorists.
 
In this talk we will however
focus on  
the second question (ii), trying to relate 
  the  observed ``perfect fluidity'' 
 with the confinement/deconfinement transition via
electric-magnetic duality arguments. 
(The reader who wants to jump to the main qualitative argument
may go directly to section \ref{sec_bottle}).

\begin{figure}[t]
  \includegraphics[width=0.6\textwidth]{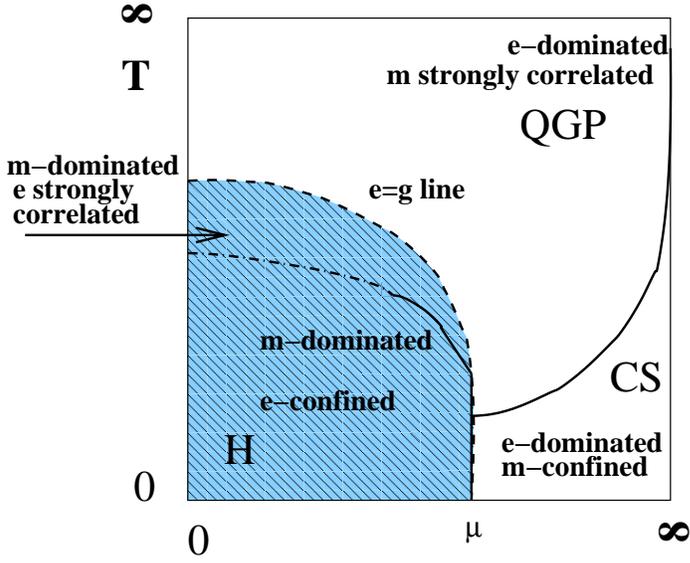}
  \vspace{0.1in}
\caption{(color online) A schematic phase diagram on
 a (``compactified'') plane of
 temperature and baryonic chemical potential $T-\mu$.
 The (blue) shaded
region shows ``magnetically dominated'' region $g<e$, which
includes the e-confined hadronic phase as well as ``postconfined''
part of the QGP domain. Light region includes ``electrically
dominated'' part of QGP and also color superconductivity (CS)
region, which has e-charged diquark condensates and therefore
obviously m-confined. The dashed line called ``e=g \,line'' is the
line of electric-magnetic equilibrium.
 The solid lines indicate true phase transitions,
while the dash-dotted line is a deconfinement cross-over line}
\label{fig_em_phasediag}
\end{figure}

\section{Electric-magnetic struggle}
\label{sec_struggle}

 Let me start by explaining what I mean by ``struggle''
in the title. A charge entering a region of space with a magnetic
 field has large chances to make fraction of a Larmor circle and
get out. Thus the Lorentz force would transfer momentum in a way
producing a pressure on a magnetic field. Therefore ``electrically
dominated'' matter have tendency to expel magnetic field into
flux tubes. We know this is the case  in superconductors
or in (e.g. Solar) plasmas.  

  According to well known
 t'Hooft-Mandelstamm scenario,  the QCD confined phase
 is a ``dual superconductor'', in which some magnetically charged
objects are Bose-condensed. By duality, now it is the electric field
 which is subdominant and thus expelled from volume into flux tubes.
From tension and also size of the ``QCD strings'', as we used 
to call them, we know that the pressure on it is quite substantial.

{\bf The picture}
proposed in \cite{Liao:2006ry} has three main ideas:\\
(i) that transition from magnetically dominated confined phase to
electrically dominated wQGP at high $T$ must proceed via
a region of approximate electric-magnetic equilibrium;\\
(ii) this must happen in the deconfined phase which however
has significant fraction of magnetic quasiparticles in a ``normal''
(uncondenced) phase;\\
(iii) in this region Dirac condition demands equality of
electric and magnetic couplings: this both $must$ be large.

A phase diagram\footnote{The arguments use continuity of the
  transition and so they are most adequate either for
pure gauge theory at $N_c=2$ or QCD with quarks in which
transition is a crossover. Large $N_c$ case involves large jump
in densities of all components which obscure the picture.
} explaining this pictorially is shown in
Fig.\ref{fig_em_phasediag}:
its blue-shaded part is 
``magnetically dominated'' region at lower $T,\mu$, while white is
``electrically dominated'' region at large $T$ and/or\footnote{Color
  superconductor at large density is electrically dominated
by diquarks while magnetic objects are confined there.} $\mu$. They are
separated by  ``E-M equilibrium'' line, which
we define by  the condition that {\em couplings}
of both interactions 
are equal\footnote{We use field theory  notations, in which
$g_e,g_m$ are electric and magnetic couplings, e/m
duality transformation is $\tau->-1/\tau$ where $\tau=\theta/2\pi+i
4\pi/g_e^2$. $g_e=g$ and $\hbar=c=1$ elsewhere.}
\be g_e^2/{4\pi\hbar c}= g_m^2/{4\pi\hbar c}= 1 \ee
The last equality follows from the
 celebrated Dirac quantization condition
\begin{equation} \label{Dirac_quantization}
\frac{g_eg_m}{4\pi\hbar c} = \frac{n}{2}
\end{equation}
with $n$ being an integer\footnote{Which we  put to n=2 because of
a presumed adjoint color representation
of relevant monopoles.}.
The ``magnetic-dominated'' low-$T$ (and low-$\mu$) region (i) can
in turn be subdivided into the $confining$ part (i-a) in which
electric field is confined into quantized flux tubes  by
magnetic condensate 
 \cite{t'Hooft-Mandelstamm}, and a new {\em ``postconfinement''}
 region (i-b) at $T_c<T<T_{E=M}$ in which the electric quasiparticles
-- quarks and gluons -- already exist but they are heavy and
still  sub-dominant in number as compared to magnetic ones.
  This picture, among many other new viewpoints, opens a completely new
  perspectives on the role of 
electric flux tubes: they do not disappear
at $T=T_c$ but are rather at their maximum.

Besides equal couplings, the equilibrium region is also
presumably characterized by comparable densities as well as masses
of both electric and magnetic quasiparticles. In QCD the issue
is complicated by the fact that E-M duality is far from perfect,
with different spins of electric (gluons and quarks) and magnetic
quasiparticles. However other theories -- especially  $\cal N$=4
supersymmetric YM -- have perfect self-duality of electric and
magnetic description: it is also conformal and has no confinement
to complicate the picture, while E and M-dominated parameter regions
do exist\footnote{Not in the phase diagram, as in this theory
  couplings are independent of $T,\mu$.}.

 If so, there should be
be uncondensed magnetic objects above $T_c$ as well. And indeed,
 lattice studies show
that electrically charged particles -- quarks and gluons --
are getting heavier as we decrease $T$ toward $T_c$,
 while monopoles gets lighter and more numerous.\\
  The magnetic screening mass,
although absent  perturbatively, 
%
is nonzero, and even $exceeds$ the electric one close to $T_c$  
(as shown e.g. by Nakamura et al \cite{Nakamura:2003pu}).
\footnote{Note that electric mass is $M_E\sim gT$  is
 larger than the magnetic
one $M_M\sim g^2T$ at high $T$. $M_E$
however vanishes at $T_c$ 
while $M_M$ smoothly grows into confined phase toward $T=0$.}
These screening masses as well as estimates of the densities of
electric and magnetic objects, 
leads to the location of {\em
E-M equilibrium} \cite{Liao:2006ry} at
 \be T_{E=M}\approx (1.2-1.5)T_c=250-300\, MeV\ee.

Seiberg-Witten solution for the \cal{N}=2 SYM is an excellent
example of how transition from
one language to another is  supposed to work, as the electric and
magnetic couplings run in the opposite directions.
And indeed, 
as one approaches the deconfinement transition
the electrically charged particles -- quarks and gluons --
are getting heavier and heavier
while the magnetic objects -- monopoles and/or
dyons -- gets lighter and more numerous.

(The stumbling difference between supersymmetric models and QCD
is of course a protection of any Higgsed vacuum by supersymmetry
in the former cases, making them all degenerate and thus stable.
We cannot discuss in detail a QCD setting: the reader may
simply imagine a generic finite-$T$ configuration with
a nonzero mean $<A_0>$, an  adjoint Higgsing leaving
$N_c-1$ U(1) massless gauge fields. These U(1)'s corresponds to
magnetic charges of the monopoles. In AdS/CFT language one may simply
considered $N_c$ branes to be placed not at exactly the same point
in the orthogonal space.)

\section{Static potentials and flux tubes}

 At $T=0$ static potential between
heavy quarks is the sum of the Coulomb and the confining 
$\sim\sigma(T=0) r$ 
potentials. At nonzero $T$ expectations of the
Wilson or Polyakov lines with
a static quark pair define the free energy potential $F(T,r)$,
which at the deconfinement point $T=T_c$
has  vanishing string tension $\sigma(T_c)=0$.

 However the $energy$
or $entropy$ separately\footnote{Related as usual by $F=E-TS$,
  $S=-\partial F/\partial T$.}
continue to have linear part  in the
``postconfinement'' domain $T_c<T< T_{E=M}$, as follows from
lattice data \cite{potentials}. 
 Quite shockingly, the
 tension term in E(T=Tc,r) is not smaller but
actually  twice larger than the zero temperature tension
  $\sigma(T=0)$! The total energy 
added to a pair till it finally breaks into two objects
is surprisingly large:
 $E(T=T_c,r\rightarrow\infty)\sim 4\, GeV$, while the corresponding
entropy as large as $S(T=T_c,r\rightarrow\infty)\sim 20$. 

Where all this huge energy and entropy comes from, at $T_c$ and in the
 ``postconfinement'' domain? Linear dependence on $r$ hints that  
it must still be a string-like object 
connecting two static charges. 
Can it be that magnetically dominant plasma admits
(metastable) electric 
flux tubes solutions?  The answer to this question
was given in our recent paper \cite{Liao:2007mj}. Indeed,
whether monopoles  are condensed or not is not so crucial:
what is important is the relation between their energy
and the repulsive potential of the flux tube. At high enough $T$
 flux tube solution disappears, and the critical condition
we found roughly does correspond to the upper limit of the
``postconfinement'' domain.  

\section{The ``magnetic bottle'' effect}
\label{sec_bottle}
\begin{figure}
\includegraphics[width=0.8\textwidth]{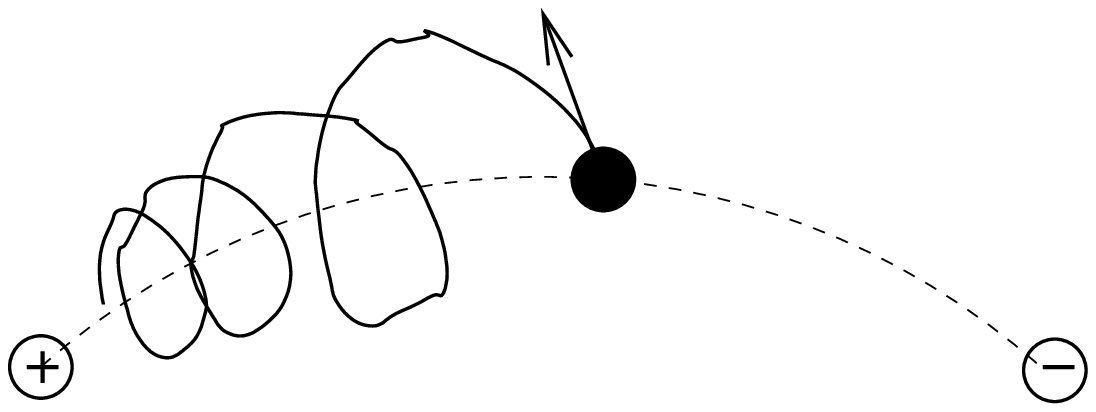}\\
\includegraphics[width=0.9\textwidth]{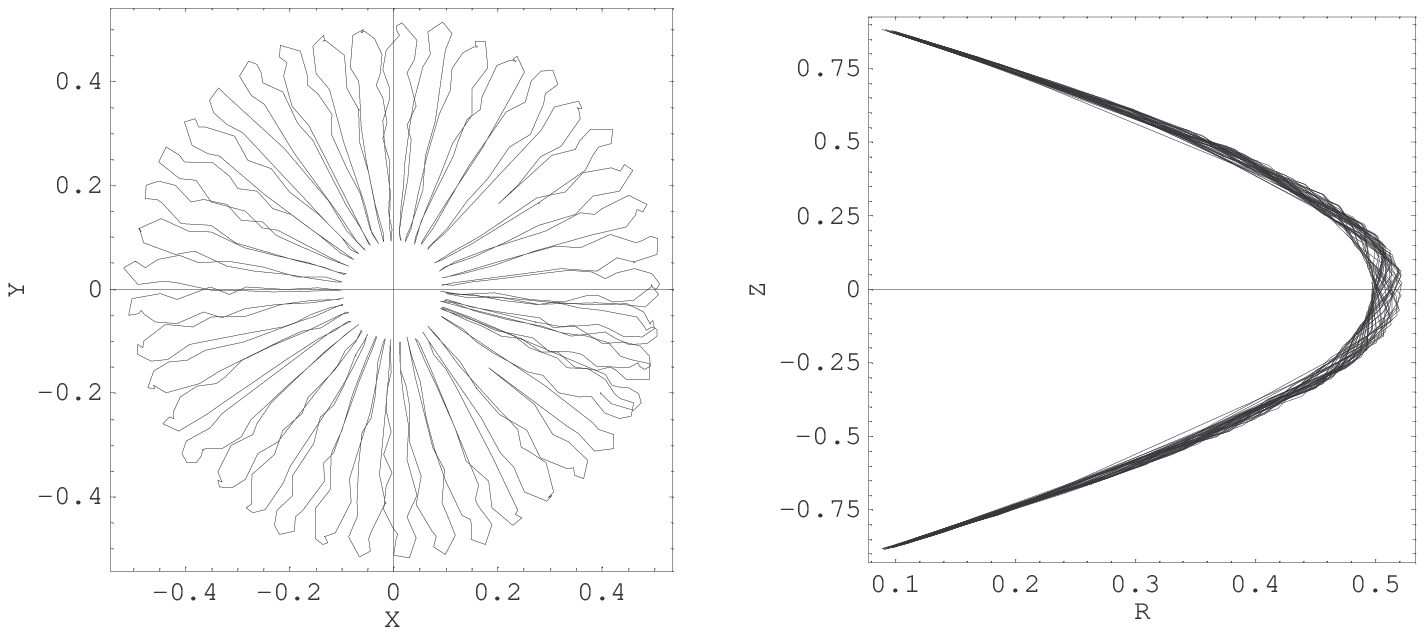}
 \caption{\label{fig_trap_1}
(a) Schematic setting of 
 a monopole moving in a static electric field of two charges
 $\pm 1$. (b) A sample of real trajectory projected on
x-y plane   (left panel) and projected on R-z plane (right panel)
(with $R=\sqrt{x^2+y^2}$ being transverse radius).}
 \end{figure}

  Before we present manybody studies of plasmas
made of electrically and magnetically charged quasiparticles,
we would like to explain the essence of the effect using the
smallest number of particles possible. The minimal
arrangement shown in Fig\ref{fig_trap_1}(a) 
 includes two static particles of one kind
(say +1 and -1 electric charges) and one dynamical particle of the 
other kind (a monopole of any sign). 

Naively one may think that with a velocity $\vec v$ (the arrow) 
looking away from both charges there are no chances of a collisions
with them, with any appreciable momentum exchange. However
due to the Lorentz force, 
    monopoles with not too large velocity rotate around
  the electric field line (dashed) while drifting along it.
  As the monopole approaches one of the charges,  the field gets 
stronger
  and the Larmor radius smaller: the motion gets confined 
  to the so called Poincare cone and collision happens. 
 Effective repulsion
(see quantum discussion below) pushes the particle away from the charge, only to repeat
  the same thing at the other charge. 

 By duality, the same will happen
for a charges surrounded by two monopoles. 
  The principle is thus the same as in famous Budker's ``magnetic
 bottle" invented precisely for ion trapping: only
  in this case the dynamical particle has electric charge
 while the static charges are substituted by  small coils. The reason a particle
 is effectively repelled from high field regions inside the coil is due to
 conservation of adiabatic invariants, and is not absolute.

In summary, in significant part of the phase
space a monopole is classically 
trapped, and cannot avoid periodical head-on 
collisions with the charges.
As a result, particle diffusion is strongly reduced,
and a plasma made  of electric and magnetic
charges can move collectively even for rather small systems.

\begin{figure}
\includegraphics[width=0.6\textwidth]{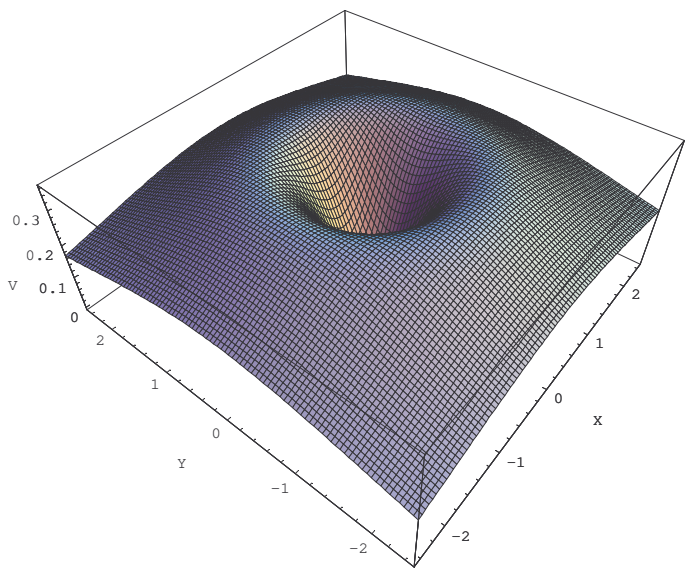}
 \caption{\label{fig_effective_pot}
 (color online) Quantum mechanical effective potential for a
 monopole in a static electric dipole field. See text for details.
}
 \end{figure}

 Addressing the same problem quantum mechanically,  one has
 the following Hamiltonian for the monopole:
\begin{equation} \label{dipole_hamiltonian}
\hat{\cal H} =\frac{(\vec p + g {\vec A_e})^2}{2m}
\end{equation}
Here $\vec A_e$ is the electric vector potential of the dipole
electric field, which can be thought of as a dual to the normal
magnetic vector potential of a magnetic dipole made of
monopole-anti-monopole. By symmetry argument we can require the
vector potential as $\vec A_e = A^{\phi}_e(\rho,z) \hat \phi$ and
the monopole wave function as $\Psi=\psi(\rho,z) e^{i f \phi}$ with
$f$ the z-angular-momentum quantum number. Then the stationary
Schroedinger equation is simplified to be
\begin{eqnarray} \label{effective_potential}
&&[\frac{\vec p_{\rho}^{\,\, 2}+ \vec p_z^{\,\, 2}}{2m} + V_{eff}
] \psi = E \psi
\\
&&V_{eff} = \frac{\hbar^2}{2m} [\frac{1}{\rho /a }
(\frac{ge}{\hbar}\frac{\rho A_e^{\phi}}{e}+f)]^2
\end{eqnarray}
To go further one has to specify a gauge (which is equivalent to
choosing some particular dual Dirac strings for the charges) so as
to explicitly write down $A_e^{\phi}$. We use the gauge which
corresponds to the situation with one Dirac string going from the
positive charge along positive $\hat z$ axis to $+ \infty$ and the
other going from the negative charge along negative $\hat z$ axis
to $- \infty$. This gives us:
\begin{equation} \label{vector_potential}
A_e^{\phi} =
-\frac{e}{\rho}[2+\frac{z-a}{\sqrt{\rho^2+(z-a)^2}}-\frac{z+a}{\sqrt{\rho^2+(z+a)^2}}]
\end{equation}
To give an idea of the effective potential we show
Fig.\ref{fig_effective_pot} where
$V_{eff}(\rho,z,f)=V_{eff}(\sqrt{x^2+y^2},z,f)$ is plotted for the
x-y plane with $z=0$ and $f=0$. From the plot we can see that
there must also be quantum states with the monopole bounded within
the potential well around the dipole for a long time before
eventually decaying away.

\section{Molecular dynamics of electric/magnetic plasma}
\label{sec_MD}
If electric and magnetic quasiparticles can be described
classically\footnote{Which is possible as soon as their masses
are large compared to $T$, which is approximately the case
close to $T_c$. It is better
satisfied  for electric than magnetic ones, however.}, we know how do
describe their ensemble, even at strong coupling. One should
$not$ use Boltzmann eqn or other simplistic transport, but use
instead   Molecular Dynamics (MD),
which means solving numerically equations of motion. It is quite
practical to do so to $\sim 1000$ particles
and to abandon periodic boundary conditions: see the details
in \cite{Liao:2006ry}.

The summary of the sQGP transport properties is shown 
 as a log-log plot of properly normalized
dimensionless (heavy quark) diffusion constant and viscosity
in Fig.\ref{fig_em_phasediag}.
 The weakly coupled (gas) regime is in the left lower corner\footnote{
It has slope one since in this case both quantities are proportional
to the same mean free path.}. 
RHIC phenomenology  (corresponding to the shaded oval) is however
quite far from the weakly coupled  regime.
The AdS/CFT line\footnote{Viscosity according
to \protect\cite{PSS} with $O(\lambda^{-3/2})$ correction, diffusion
constant from \protect\cite{Casalderrey-Solana:2006rq}.}
 on the other hand, neatly
 crosses through it, leading to optimistic conclusions.
 However our MD results, especially for 50-50\% 
electric-magnetic plasma, are also in the right ballpark.
We will discuss it a bit more in conclusions.

\begin{figure}[t]
 \includegraphics[width=.9\textwidth]{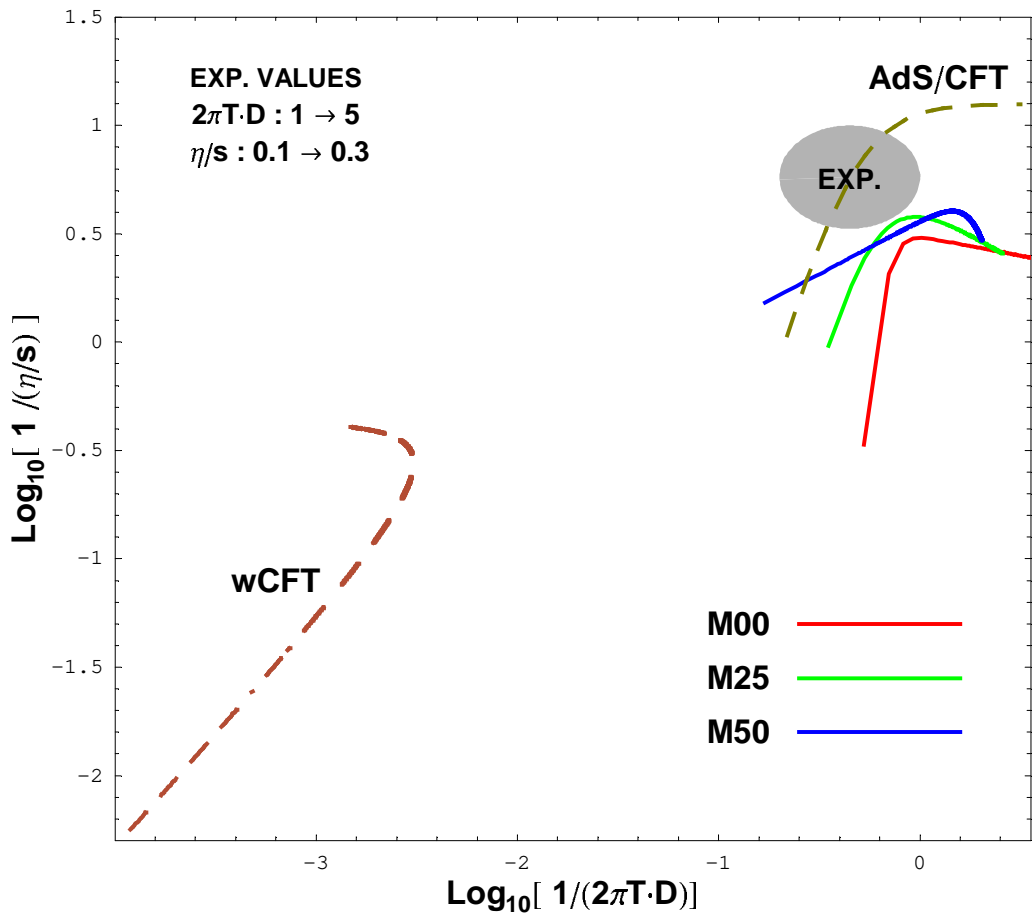}
  \vspace{0.1in}
\caption{
 Plots of $Log[1/(\eta /s)]$ v.s. $Log[1/(2\pi T D)]$ including
results from our MD simulations, the Ads/CFT calculations, the weakly coupled
CFT calculations, as compared with experimental values. M00,M25,M50
mean 0,25 and 50\% of monopoles in plasma. 
\label{fig_trasport}}
\end{figure}

\section{Are there baryons above $T_c$?}
\label{sec_baryons}
 A talk at GSI cannot be without a special discussion of the FAIR
physics domain, a dense baryonic matter. One issue
I would like to discuss briefly at the end is put in the title of this
section. Indeed, confinement is not
necessary for binding, and if some mesons survive the phase transition
into the plasma phase, why not baryons?

\begin{figure}
\begin{center}
\includegraphics[width=10cm]{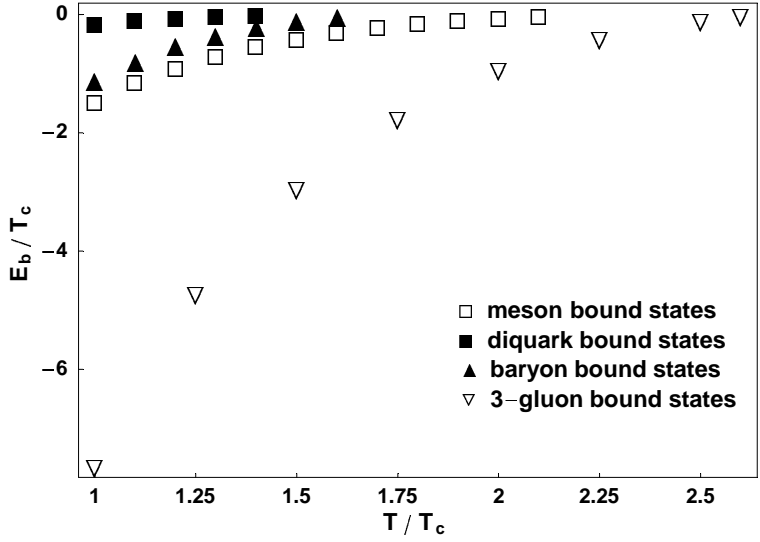}
\end{center}
 \caption{\label{fig_boundstates}
Dependence of various states' binding energy on the temperature.
The units are in $T_c$.}
 \end{figure}

The first thing to do is of course simply to use quantum mechanics
and see in which domain
 the  potentials between the QGP quasiparticles
used for mesons would bind baryons as well. This was done some time
ago in \cite{Liao:2005hj},
with results summarized in Fig.\ref{fig_boundstates}. One can see
that while e.g. diquarks have negligible binding already at $T_c$,
the s-wave baryons -- N,$\Delta$... -- seem to survive up to 1.6$Tc$.

The next step came from Bielefeld-UK collaboration \cite{Allton:2005gk} which
calculated susceptibilities (up to sixth derivative over baryonic
chemical potential $\mu_B$ and/or isospin one $\mu_I$)
and thus provided intriguing information on the QGP thermodynamics
at nonzero $\mu$. Unexpected
``peaks'' in the 4-th and ``wiggles' in the 6-th were found in their 
$T$-dependence. Our analysis of these data \cite{Liao:2005pa} 
concluded
that they can be naturally explained in a scenarios with surviving
baryons, provided baryonic masses $grow$ above $T_c$. Ratti,Roessner
and Weise   \cite{Ratti:2007jf} have provided another impressive
description of the same data in the PNJL model. 
In fact their conclusions are not that different from ours: indeed,
the main role of their nonzero Polyakov loop is to suppress quarks and
diquarks in favor of zero-triality $qqq$ states. Although the latter
are not treated as such, they are effectively the
 baryons we used in our work \cite{Liao:2005pa}.  
 
In conclusion:  we have reasons to believe that sQGP at FAIR
will be mostly made of baryons, with a relatively small admixture
of ``ionized'' quarks. Direct lattice investigations of 
baryonic correlation functions at finite $\mu$ will tell us more about
their properties. 

 Fig.\ref{fig_boundstates} shows that the most robust
baryon-like objects we found from binding calculations is
made of three gluons. In fact  $N_c$ (=3)  
adjoint objects bind a bit differently from quarks: instead of
a junction to which all colors are attracted, there is a closed chain
involving all colors cyclically closed to itself, e.g.
$(\overline{red}- blue)  (\overline{blue}- green)(\overline{green}-
red)$.

 This talk, emphasizing importance of the electric-magnetic duality,
 should naturally end with a question of whether $adjoint$ magnetic
quasiparticles -- monopoles and dyons -- can or cannot form similar
 bound states? Not only so, but in fact these ``dyonic baryons''
(in which $N_c$ dyons are attracted to
each other  pair-vise, both electrically and magnetically) are already 
well known:  
 in fact they are the finite-$T$ instantons -- calorons --
see \cite{Kraan:1998kp}. On the lattice they have been seen and
studied in multiple works\footnote{
In fact chains up to 6 dyons with alternating colors have been
seen, (M.Ilgenfritz, private communications).}. 
   The same construction can be obtained via a very interesting AdS/CFT
picture with wrapping branes \cite{Lee:1997vp}. There is no doubt 
we will have a zoo full of  exotic animals in the FAIR domain:
the question is whether we will be smart enough to trace them
down.

\section{Conclusions}
Theory of strongly coupled QGP has developed into two different
brunches: one, based on AdS/CFT emphasizes quasi-$conformal$ behavior
and may be applicable at $T>1.5Tc$ or so; another emphasizes
proximity to deconfinement and changes in electric-magnetic 
equilibrium.
As our brief ``transport summary'' in Fig.\ref{fig_trasport} indicate 
that at the moment $both$ may claim
to be an explanation of RHIC data. 

Thus we face a dilemma: Are these two explanations
 (i) deeply related and basically tell us
the same story in two different languages, or (ii)
there is no deep
relation between them and we just see a mere coincidence
of some transport coefficients? 
At the moment our understanding of AdS/CFT plasma at the
microscopic level is too poor to tell. 

The $\cal N$=4 SYM theory involved in AdS/CFT correspondence has, by 
itself,  an electric-magnetic duality, in a form
even much more perfect than QCD. It has Higgsed moduli spaces 
and magnetically charged supermultiplet, which happen
to be the same as the original gauge-fermion-scalar one
of the electric sector. However, in order to see electric-magnetic
transitions we speak about one has to give up the
large number of colors $N_c\rightarrow \infty$  
limit, which is at this moment
always used in practical AdS/CFT applications\footnote{It
obscures the issue, making both
electric and magnetic interactions infinitely strong.}.   

 Perhaps we will see resolution of this dilemma
 first from experiment. If the 
LHC collisions turn out to
  be similar to RHIC ones, with hydro-predicted
elliptic flow, it would mean we have sQGP in a quasi-conformal domain.
If not and sQGP is only there near $T_c$, the electric-magnetic
struggle would be the one left.

\end{document}